\documentstyle[12pt]{article}
\begin{document}
\begin{center}
{\large \bf Small $x$ physics in deep inelastic lepton scattering}
 \footnote{A lecture given at the XXXVth Jubillee Cracow School of
Theoretical Physics, June 1995, Zakopane, Poland}\\
\bigskip
J. Kwieci\'nski \\
Department of Physics\\
University of Durham\\
Durham, UK\\
and\\
Department of Theoretical Physics\\
H. Niewodnicza\'nski Institute of Nuclear Physics\\
Krak\'ow, Poland. \footnote{Permanent address}\\
\end{center}
\vspace{40mm}
\begin{abstract}
The perturbative QCD predictions concerning  deep inelastic scattering
at low $x$ are summarized.   The theoretical framework based
on the leading log $1/x$ resummation and $k_t$ factorization theorem
is described. The role of studying final states in deep inelastic
scattering for revealing the details of the underlying dynamics at low
$x$ is emphasised and some dedicated measurements, like deep inelastic
scattering accompanied by an energetic jet, the measurement of the transverse
energy flow and deep inelastic diffraction,  are briefly
discussed.
\end{abstract}
\newpage
\section*{1. Introduction}
It has been known for quite some time \cite{GLR,BCKK} that perturbative QCD
predicts several new phenomena to occur when the parameter $x$ specifying
the longitudinal momentum fraction of a hadron carried by a parton
(i.e. by a quark or by a gluon)
becomes very small. The main expectation was that the gluon and quark densities
should  strongly grow in this limit eventually leading to the parton
saturation effects \cite{GLR,BCKK,ADM1,JK1}.
This increase of parton distributions
implies similar increase of the deep inelastic lepton - proton
 scattering structure function $F_2$
with the decreasing Bjorken parameter $x$ \cite{AKMS}
and the experimental data from HERA
are consistent with this prediction \cite{H1,ZEUS}.
The Bjorken parameter $x$
is as usual defined  as $x=Q^2/(2pq)$ where $p$ is the proton four momentum,
$q$
the four momentum transfer between the leptons and $Q^2=-q^2$ (see Fig.1).\\

The growth of structure funcctions with decreasing parameter $x$ is
much stronger than that
which would follow from the expectations based on the "soft" pomeron
exchange mechanism with the soft pomeron intercept
$\alpha_{soft} \approx 1.08$
as determined from the phenomenological analysis of total hadronic
and real photoproduction cross-sections \cite{DOLA}.\\

Small $x$ behaviour of structure functions
for fixed $Q^2$ reflects the high energy behaviour of the
virtual Compton scattering total cross-section with increasing
total CM energy
squared $W^2$ since $W^2=Q^2(1/x-1)$. The Regge pole exchange picture
\cite{PC} is
therefore quite appropriate for the theoretical description
of this behaviour.  The high energy behaviour which follows
from perturbative QCD is often refered to as being related to the "hard"
pomeron in contrast to the soft pomeron describing the high energy
behaviour of hadronic and photoproduction cross-sections.\\

The  inelastic lepton - nucleon scattering is related through the one photon
exchange mechanism illustrated in Fig. 1
to the  Compton scattering of virtual photons.
In Fig. 2 we summarize the present experimental situation on the
variation of the total virtual Compton scattering cross-section
with total CM  energy $W$ for different photon virtualities $Q^2$  which range
from the real photoproduction ($Q^2=0$) to the deep inelastic region \cite
{ALEVY}.
The change of high energy behaviour with the scale $Q^2$ is evidently
present in the data.\\

The purpose of this lecture is to summarize briefly the QCD expectations
for the deep inelastic scattering at low $x$.  In the next section
we recall the parton model  and  the Regge pole model expectations
for the description of the small $x$ behaviour of deep inelastic
scattering structure functions.   Sec. 3 is devoted to the
description of the QCD improved parton model based on the Altarelli-Parisi
evolution equations  and the collinear factorization \cite{AP,REYA}.
We  discuss
the small $x$ behaviour of structure functions which follows from
this formalism.  In sec. 4 we summarize the results
based on the leading
ln$1/x$ resummation which is provided by the Balitzkij, Fadin, Kuraev, Lipatov
(BFKL) equation and $k_t$ factorization.  We also briefly discuss the
Catani, Ciafaloni, Fiorani
 Marchesini (CCFM) equation based on the angular ordering of gluon emission
\cite{CIAF,CCFM}.
This equation embodies the BFKL $ln1/x$ resummation and the conventional
QCD evolution in the regions of small and large values of $x$ respectively.
Sec. 5 is devoted to a brief discussion of the dedicated measurements
of the hadronic final state
in deep inelastic scattering for revealing the dynamical QCD
expectations at low $x$.  Sec. 6 contains a brief summary and conclusions.
\section*{2. Parton model description of deep
inelastic lepton scattering}
The  inelastic lepton - nucleon scattering is related through the one photon
exchange mechanism and through the optical theorem to the imaginary part
of the forward Compton scattering amplitude of virtual photons (see Fig. 1).
The latter is defined by the tensor   $W^{\mu \nu}$ \cite{HM,RGR}:
$$
W^{\mu \nu}(p,q) = {F_1(x,Q^2)\over M} \left(-g_{\mu \nu} +
{q^{\mu}q^{\nu}\over q^2}\right)+
$$
\begin{equation}
{F_2(x,Q^2)\over M(pq)}\left(p^{\mu} -{pq\over q^2}q^{\mu}\right)
\left(p^{\nu} -{pq\over q^2}q^{\nu}\right)
\label{wmunu}
\end{equation}
In this equation $p$ denotes the four momentum of the nucleon,
$Q^2=-q^2$ where $q$ is the four momentum
transfer between the leptons, $x=Q^2/(2pq)$ the Bjorken scaling variable
and $M$ is the
nucleon mass.  The  functions $F_{1,2}(x,Q^2)$ are the nucleon
structure functions.  These structure functions are directly related
to the total cross-sections $\sigma_L$ and $\sigma_T$ corresponding to
the longitudinaly and transversely polarized virtual photons respectively:
\begin{equation}
F_2={Q^2\over 4 \pi^2 \alpha}(\sigma_T + \sigma_L)
\label{f2}
\end{equation}
\begin{equation}
F_L=F_2-2xF_1={Q^2\over 4 \pi^2 \alpha}\sigma_L
\label{fl}
\end{equation}
The differential cross-section describing  inelastic lepton scattering
is related in the following way to the structure functions $F_2$ and $F_1$:
\begin{equation}
{d^2\sigma(x,Q^2)\over dx dQ^2}
={4 \pi \alpha^2\over Q^4}\left [ \left (1 - y \right)
{F_2(x,Q^2)\over x} + y^2F_1(x,Q^2) \right]
\label{dsigma}
\end{equation}
where $y=pq/p_lp$ with $p_l$ denoting the four-momentum of the
incident lepton.\\

 The deep inelastic regime is defined as the region
where both $Q^2$ and $2pq$ are large and their ratio, $x$, is fixed.
It is assumed that in this region the deep inelastic scattering
reflects the (elastic) lepton scattering on (point-like) quarks and
antiquarks, as illustrated by the "hand-bag" diagram of Fig. 3,
with the virtuality $k^2$ of the quark (antiquark)
being limited. The "hand- bag" diagram for virtual Compton
scattering together with the limitation of the
virtuality $k^2$ form the basis of the (covariant) parton model which leads
to Bjorken scaling (i.e. $F_2(x,Q^2) \rightarrow F_2(x)$ ) \cite{CVPM}.
In the infinite momentum frame the Bjorken variable $x$
acquires the meaning of the momentum
fraction of the parent nucleon carried by a probed quark (antiquark).
For spin $1/2$ partons the structure function $F_L$ vanishes
in the deep inelastic region- this being a straightforward consequence
of the limitation of the quark transverse momentum \cite{HM,RGR}.\\

In the parton model the structure function $F_2$ becomes directly
related to the quark and antiquark distributions $q_i(x)$ and $\bar q_i(x)$
in a nucleon
\begin{equation}
F_2=x\sum_i e_i^2[q_i(x)+ \bar q_i(x)]
\label{f2pm}
\end{equation}
where $e_i$ is the charge of the quark carrying flavour "$i$".
At small $x$ (i.e. for $2pq >> Q^2$) one can use the Regge pole model for
parametrizing the high energy behaviour of the  total cross-sections
$\sigma_{T,L}$ and obtain, using eq. (\ref{f2}),  the following
parametrization of the structure function $F_2(x,Q^2)$:
\begin{equation}
F_2=\sum_i \tilde \beta_i x^{1-\alpha_i(0)}.
\label{reggef}
\end{equation}
The relevant reggeons are
those which can couple to two (virtual) photons.  The (singlet) part
of the structure function $F_2$ is controlled at small $x$ by
the pomeron exchange, while the non-singlet part $F_2^{NS}=
F_2^p-F_2^n$ by the $A_2$ reggeon. From equations (\ref{reggef})
and (\ref{f2pm}) one gets
the corresponding Regge behaviour of
quark (antiquark) distributions.

\section*{3. Small $x$ limit of parton distributions in the QCD -  improved
parton model}

The Bjorken scaling (i.e. independence of the structure functions
of $Q^2$) is violated by the elementary QCD interactions and
the parton (i.e. the quark, antiquark and gluon
distributions) acquire  $Q^2$ dependence. Their change  with $Q^2$
is described by the Altarelli-Parisi evolution equations
\cite{AP,REYA}.  In the leading
$ln(Q^2)$ approximation  which resums the leading powers of $\alpha_sln(Q^2)$
the eq.  (\ref{f2pm}) still holds, although the
quark (and antiquark) distributions
are now scale dependent. This scale dependence comes from
the fact that the quark (antiquark) virtuality $k^2$ in the
"hand-bag" diagram of Fig. 3 is no longer limited
as   in the "naive" parton model.
This lack of limitation of $k^2$ is the result of the (point-like)
elementary QCD
interactions.  The dominant region from which the scaling violations come
 is now $k_0^2<<k^2<<Q^2$, where $k_0^2$ is some infrared cut-of.
  Beyond the leading $ln(Q^2)$ approximation the eq.(\ref{f2pm}) acquires
$O(\alpha_s)$ corrections which are however (by definition) absent in the
so called deep inelastic (DIS) scheme. \\

At small $x$ the dominant role is played by the gluons and so
for simplicity we shall limit ourselves
to the following system of the evolution equations:
\begin{equation}
Q^2{dq(x,Q^2)\over d Q^2}=\int_x^1{dz\over z}P_{qg}
(z,\alpha_s(Q^2))g({x\over z},Q^2)
\label{apq}
\end{equation}
\begin{equation}
Q^2{dg(x,Q^2)\over d Q^2}=\int_x^1{dz\over z}P_{gg}
(z,\alpha_s(Q^2))g({x\over z},Q^2)
\label{apg}
\end{equation}
where $q(x,Q^2)$ and $g(x,Q^2)$ are the scale dependent quark and gluon
distributions respectively.
The splitting  functions $P_{ij}(z,\alpha_s(Q^2))$ can be expanded
in the perturbative series of the QCD coupling $\alpha_s(Q^2)$.
In  leading order
\begin{equation}
P_{ij}(z,\alpha_s(Q^2)) = {\alpha_s(Q^2)\over 2 \pi}
P_{ij}^{(0)}(z)
\label{pij}
\end{equation}
 In an axial gauge the leading order evolution equations
sum ladder diagrams with ordered longitudinal and strongly ordered
transverse momenta along the chain.\\

The  evolution equations can be solved in a
closed form for the
moment functions $\bar h(\omega,Q^2)$  of the parton distributions $h(x,Q^2)$
\begin{equation}
\bar h(\omega,Q^2)= \int_0^1{dx\over x}x^{\omega} xh(x,Q^2)
\label{hmom}
\end{equation}
where $h(x,Q^2)$ denotes the gluon or quark (antiquark) distribution.
 The distribution
$h(x,Q^2)$ is related by the inverse transform  to the moment function
\begin{equation}
h(x,Q^2)={1\over 2 \pi i} \int_{c-i\infty}^{c+i\infty}d\omega
x^{\omega-1}\bar h(\omega,Q^2)
\label{hinv}
\end{equation}
where the integration contour is located to the right of the leading
(i.e. rightmost) singularity of the moment function $\bar h(\omega,Q^2)$
in the $\omega$ complex plane.
The use of moments is therefore
useful for understanding the small $x$ behaviour which is controlled
by the leading singularity of the moment functions  in the $\omega$ plane.\\
The solution
of the counterpart of the evolution equation (\ref{apg})
for the moment function $\bar g(\omega,Q^2)$
\begin{equation}
\bar g(\omega,Q^2)=\int_0^1{dx\over x}x^{\omega} xg(x,Q^2)
\label{momg}
\end{equation}
 is of the form:
\begin{equation}
\bar g(\omega,Q^2)=\bar g(\omega,Q_0^2)exp\left[\int_{Q_0^2}^{Q^2}{dq^2\over
q^2}
\gamma_{gg}(\omega,\alpha_s(q^2))\right ]
\label{gev}
\end{equation}
and, similarly  the quark distributions "driven by the gluon" are given by:
\begin{equation}
Q^2{d\bar q(\omega,Q^2)\over d Q^2}=
\gamma_{qg}(\omega, \alpha_s(Q^2)) \bar g(\omega,Q_0^2)
exp\left[\int_{Q_0^2}^{Q^2}{dq^2\over q^2}
\gamma_{gg}(\omega,\alpha_s(q^2))\right ].
\label{qdbg}
\end{equation}
The anomalous dimensions $\gamma_{ij}(\omega,\alpha_s(Q^2))
$ are moments of the splitting functions $P_{ij}(z,\alpha_s(q^2))$
\begin{equation}
\gamma_{ij}(\omega,\alpha_s(Q^2))=\int_0^1{dz\over z}z^{\omega} z
P_{ij}(z,\alpha_s(Q^2)).
\label{spm}
\end{equation}
 In the leading order
\begin{equation}
\int_{Q_0^2}^{Q^2}{dq^2\over q^2}
\gamma_{gg}(\omega,\alpha_s(q^2))=\gamma_{gg}^{(0)}(\omega)
\xi(Q^2,Q_0^2)
\label{lo}
\end{equation}
where
\begin{equation}
\xi(Q^2,Q_0^2)=\int_{Q_0^2}^{Q^2}{dq^2\over q^2}{\alpha_s(q^2)\over 2\pi}
\sim log \left({log({Q^2\over \Lambda^2})\over log({Q_0^2\over \Lambda^2})
}\right)
\label{xilo}
\end{equation}
and $\gamma_{gg}^{(0)}(\omega)$ is the moment of $P_{gg}(z)$. \\

The singularities of the moment function $\bar g(\omega,Q^2)$ in the $\omega$
plane are present both in the anomalous dimension
$\gamma_{gg}(\omega,\alpha_s(q^2))$ as well as
in  the moment of the input nonperturbative gluon distribution
$\bar g(\omega, Q_0^2)$, see (\ref{gev}). The small $x$ behaviour of the gluon
 distributions is controlled
by the leading  singularity. The same singularity should also control
the small $x$ behaviour of the (sea) quark distributions through the
$g \rightarrow q\bar q$ transitions which, within the QCD evolution
formalism, are described by eq.(\ref{qdbg}).
In  leading order, the anomalous dimension
$\gamma_{gg}(\omega,\alpha_s(q^2))$ has a pole at $\omega=0$
since $P^{0}_{gg}(z) \sim 6/z$ at small $z$.   It leads to the
essential singularity of the moment function  $\bar g(\omega,Q^2)$
at $\omega=0$ (see eq. (\ref{gev})).
 This  esssential
singularity   remains the
leading one  provided that  the starting gluon
distribution $x\bar g(x,Q_0^2)$ behaves (at most) as a constant
at small $x$.  One gets then
 the following "double logarithmic behaviour"
for the gluon  distribution at small $x$:
\begin{equation}
xg(x,Q^2) \sim exp\left(2\sqrt{6\xi(Q^2,Q_0^2)ln({1\over x})}\right)
\label{dl}
\end{equation}
with similar behaviour for the sea quark distributions.
If a more singular behaviour is taken for the input $xg(x,Q_0^2)$
then it remains stable against
 leading order QCD evolution in $Q^2$.

\section*{4. The BFKL pomeron and QCD predictions for the small $x$
behaviour of
the deep inelastic scattering structure functions}
The QCD improved parton model in which the splitting functions $P_{ij}(z)$
are computed at fixed order of their perturbative expansions is incomplete
in the small $x$ region.  In this region the perturbative terms generate
powers of $ln(1/x)$ and we should, at least, resum
 the powers of $\alpha_s ln(1/x)$ i.e. to consider the leading $ln(1/x)$
approximation. The basic dynamical quantity now is the
unintegrated gluon distribution
$f(x,Q_t^2)$ where $x$ denotes the momentum fraction
of a parent hadron carried by a gluon and $Q_t$  its transverse
momentum.  The unintegrated distribution $f(x,Q_t^2)$
is related in the following way to the more familiar scale dependent
gluon distribution $g(x,Q^2)$:
\begin{equation}
xg(x,Q^2)=\int^{Q^2} {dQ_t^2\over Q_t^2} f(x,Q_t^2).
\label{intg}
\end{equation}
In the leading $ln(1/x)$  approximation the unintegrated
distribution $f(x,Q_t^2)$ satisfies
the BFKL equation \cite{BFKL} which has the following form:
$$
f(x,Q_t^2)=f^0(x,Q_t^2)+
$$
\begin{equation}
\bar \alpha_s \int_x^1{dx^{\prime}\over
x^{\prime}} \int {d^2 q\over \pi q^2}
\left[{Q_t^2 \over (\mbox{\boldmath $q$}+
\mbox{\boldmath $Q_t$})^2}
f(x^{\prime},(\mbox{\boldmath $q$}+
\mbox{\boldmath $Q_t$})^2)-f(x^{\prime},Q_t^2)\Theta(Q_t^2-q^2)\right]
\label{bfkl}
\end{equation}
where
\begin{equation}
\bar \alpha_s={3\alpha_s\over \pi}
\label{alphab}
\end{equation}
The first and the second
terms  on the right hand side of  eq (\ref{bfkl}) correspond
to  real gluon emission with $q$ being the transverse
momentum of the emitted gluon, and to the virtual corrections respectively.
$f^0(x,Q_t^2)$ is a suitably defined inhomogeneous term.\\

After resumming the virtual corrections and "unresolvable"  gluon
emissions ($q^2 < \mu^2$)  where $\mu$ is the resolution
defining the "resolvable" radiation,  equation (\ref {bfkl})
can be rearranged into the following "folded" form:
$$
f(x,Q_t^2)=\hat f^0(x,Q_t^2)+
$$
\begin{equation}
 \bar \alpha_s
\int_x^1{dx^{\prime}\over
x^{\prime}} \int {d^2 q\over \pi q^2} \Theta
(q^2-\mu^2)\Delta_R({x\over x^{\prime}},Q_t^2)
{Q_t^2 \over (\mbox{\boldmath $q$}+
\mbox{\boldmath $Q_t$})^2}
f(x^{\prime},(\mbox{\boldmath $q$}+
\mbox{\boldmath $Q_t$})^2) +O(\mu^2/Q_t^2)
\label{bfklr}
\end{equation}
where $\Delta_R$ which screens the $1/z$ singularity is given by:
\begin{equation}
\Delta_R(z,Q_t^2)=z^{\bar \alpha_s ln (Q_t^2/ \mu^2)}=
exp\left(-\bar \alpha_s\int_z^1{dz^{\prime}
\over z^{\prime}}\int_{\mu^2}^{Q_t^2}{dq^2\over q^2}\right)
\label{deltar}
\end{equation}
and
\begin{equation}
\hat f^0(x,Q_t^2)= \int_x^1{dx^{\prime}\over
x^{\prime}} \Delta_R({x\over x^{\prime}},Q_t^2)
{df^0(x^{\prime},Q^2_t)\over dln(1/x^{\prime})}
\label{hatf0}
\end{equation}
 Equation (\ref{bfklr}) sums  the ladder diagrams (see Fig. 4)
with the reggeized gluon exchange along the chain  with the gluon
trajectory $\alpha_G(Q_t^2) = 1-{\bar \alpha_s\over 2} ln(Q^2_t/\mu^2)$.\\

For the fixed coupling case  eq. (\ref{bfkl}) can be solved
analytically and the leading behaviour of its solution
at small $x$ is given by the
following expression:
\begin{equation}
f(x,Q_t^2) \sim (Q_t^2)^{{1\over 2}} {x^{-\lambda_{BFKL}}\over
\sqrt{ln({1\over x})}} exp\left(-{ln^2(Q_t^2/\bar Q^2)\over 2 \lambda^"
ln(1/x)} \right)
\label{bfkls}
\end{equation}
with
\begin{equation}
\lambda_{BFKL}=4 ln(2) \bar \alpha_s
\label{pombfkl}
\end{equation}
\begin{equation}
\lambda^"=\bar \alpha_s 28 \zeta(3)
\label{diff}
\end{equation}
where the Riemann zeta function $\zeta(3) \approx 1.202$.  The
parameter $\bar Q$ is of nonperturbative origin.\\

The quantity $1+ \lambda_{BFKL}$ is equal to the intercept of the so -
called BFKL pomeron. Its potentially large magnitude ($\sim 1.5$)
should be contrasted with the intercept $\alpha_{soft} \approx 1.08$
of the (effective) "soft" pomeron which has been determined
from the phenomenological analysis of the high energy behaviour
of hadronic and photoproduction total cross-sections \cite{DOLA}.\\

In practice one introduces the running coupling $\bar \alpha_s(Q_t^2)$
in the BFKL equation (\ref{bfkl}). This requires introduction of the infrared
cut-off that would prevent entering the infrared region where the
coupling becomes large. The effective intercept $\lambda_{BFKL}$
found by numerically solving the equation depends weakly
on the magnitude of this cut-off \cite{KMS2}.\\

The solution (\ref{bfkls}) of the BFKL equation is obtained most
directly by solving  the corresponding equation for the moment
function
\begin{equation}
\bar f(\omega, Q_t^2)= \int_0^1 {dx\over x} x^{\omega} f(x,q_t^2)
\label{momf}
\end{equation}
$$
\bar f(\omega,Q_t^2)=\bar f^0(\omega,Q_t^2)+
$$
\begin{equation}
{\bar \alpha_s \over \omega}  \int {d^2 q\over \pi q^2}
\left[{Q_t^2 \over (\mbox{\boldmath $q$}+
\mbox{\boldmath $Q_t$})^2}
\bar f(\omega,(\mbox{\boldmath $q$}+
\mbox{\boldmath $Q_t$})^2)-\bar f(\omega,Q_t^2)\Theta(Q_t^2-q^2)\right]
\label{bfklm}
\end{equation}
This equation can be diagonalised by a Mellin transform and its solution
for the Mellin transform $\tilde f(\omega, \gamma)$ of the moment
function $\bar f(\omega,Q_t^2)$ is:
\begin{equation}
\tilde f(\omega, \gamma)= {\tilde f^0(\omega, \gamma)\over
1-{\bar \alpha_s \over \omega} \tilde K(\gamma)}
\label{bfklsg}
\end{equation}
where
\begin{equation}
\tilde K(\gamma)=2\Psi(1)-\Psi(\gamma)-\Psi(1-\gamma)
\label{kg}
\end{equation}
is the Mellin transform of the kernel of eq. (\ref{bfklm}). The function
$\Psi(z)$ is the logarithmic derivative of the Euler $\Gamma$ function.
The Mellin transform $\tilde f(\omega, \gamma)$ is defined as below:
\begin{equation}
\tilde f(\omega, \gamma)=\int_0^{\infty}{dQ_t^2\over Q_t^2}
(Q_t^2)^{-\gamma}\bar f(\omega,Q_t^2)
\label{mt}
\end{equation}
and hence the function $\bar f(\omega,Q_t^2)$ is related to
$\tilde f(\omega, \gamma)$ through the inverse Mellin transform
\begin{equation}
\bar f(\omega, Q_t^2)= {1\over 2 \pi i} \int_{1/2 - i \infty}^{1/2 +i \infty}
d\gamma (Q_t^2)^{\gamma}\tilde f(\omega, \gamma)
\label{imt}
\end{equation}

The poles of $\tilde f(\omega, \gamma)$ in the $\gamma$ plane define the
anomalous dimensions of the
moment function $\bar f(\omega, Q_t^2)$ \cite{JAR}.
The (leading twist) anomalous
dimension $ \gamma_{gg}(\omega,\bar \alpha_s)$ of
$\bar f(\omega,Q_t^2)$ gives  the following behaviour of
$\bar f(\omega,Q_t^2)$
at large $Q_t^2$
\begin{equation}
\bar f(\omega,Q_t^2) =
\tilde f^0(\omega,\gamma=\gamma_{gg}(\omega,\bar \alpha_s))
\gamma_{gg}(\omega,\bar \alpha_s)
R(\alpha_s,\omega)
(Q_t^2)^{ \gamma_{gg}(\omega,\bar \alpha_s)}
\label{adbeh}
\end{equation}
where
\begin{equation}
R(\alpha_s,\omega) =-\left[  {\bar \alpha_s \over \omega}
\gamma_{gg}(\omega,\bar \alpha_s){d \tilde K(\gamma)\over
d\gamma} |_{\gamma= \gamma_{gg}(\omega,\bar \alpha_s)}\right]^{-1}.
\label{r}
\end{equation}
The anomalous dimension will also, of course, control
the large $Q^2$ behaviour of the moment function
$\bar g(\omega,Q^2)$ of the integrated gluon distribution
\begin{equation}
 \bar g(\omega,Q^2)=\int_0^{Q^2}{dQ_t^2\over Q_t^2}f(\omega,Q_t^2)
\label{gintm}
\end {equation}
which has the  following form:
\begin{equation}
\bar g(\omega,Q^2)=R(\alpha_s,\omega) \bar g^0(\omega)\left
({Q^2\over Q_0^2}\right)^{ \gamma_{gg}(\omega,\bar \alpha_s)}
\label{intgm}
\end{equation}
where we have introduced the moment function of the input distribution
\begin{equation}
\bar g^0(\omega)=\tilde f^0(\omega,\gamma=\gamma_{gg}(\omega,\bar \alpha_s))
(Q_0^2)^
{ \gamma_{gg}(\omega,\bar \alpha_s)}.
\label{g0}
\end{equation}
Equations (\ref{intgm})and (\ref{g0}) follow directly from  equations
(\ref{adbeh}), (\ref{r}) and (\ref{gintm}).
It may be seen from eq. (\ref{intgm}) that the BFKL singularity
affects through
the factor $R$
the "starting" gluon distribution at $Q^2=Q_0^2$ \cite{CIAFKT}.\\

It follows from eq.(\ref{kg}) that the anomalous dimension $\gamma_{gg}(
\omega,\bar \alpha_s))$
is the solution of the following equation:
\begin{equation}
{\bar \alpha_s\over \omega} \tilde K( \gamma_{gg}(\omega,\bar \alpha_s))=1.
\label{adeq}
\end{equation}
It is a function of only one variable ${\bar \alpha_s \over \omega}$
i.e. $ \gamma_{gg}(\omega,\bar \alpha_s) \rightarrow
 \gamma_{gg}({\bar \alpha_s\over \omega})$.
The solution of  eq. (\ref{adeq}) makes it possible
to obtain the anomalous dimension
$ \gamma_{gg}({\bar \alpha_s\over \omega})$
as a power series of ${\bar \alpha_s\over \omega}$ \cite{JAR}
\begin{equation}
 \gamma_{gg}({\bar \alpha_s\over \omega})= \sum_{n=1}^{\infty}
 c_n \left (
{\bar \alpha_s \over \omega}\right)^n
\label{adexp}
\end{equation}
This power series corresponds to the leading $ln(1/z)$
expansion of the splitting function $P_{gg}(z,\alpha_s)$
\begin{equation}
zP_{gg}(z,\alpha_s)=\bar \alpha_s
\sum_{n=1}^{\infty} c_n {(\bar \alpha_s ln(1/z))^{n-1}\over (n-1)!}
\label{exppgg}
\end{equation}
which controls the
evolution of the gluon distribution.
\\

The exponent $\lambda_{BFKL}$ controlling the small $x$ behaviour of
$f(x,Q_t^2)$ is
\begin{equation}
\lambda_{BFKL}=\bar \alpha_s \tilde K(1/2)
\label{lbfkl}
\end{equation}
The  anomalous dimension has a branch point singularity at
$\omega = \lambda_{BFKL}$.  We also have
\begin{equation}
 \gamma_{gg}(\omega=\lambda_{BFKL})={1\over 2}
\label{onehalf}
\end{equation}

The following properties of the solution of the BFKL equation
summarized in the formula (\ref{bfkls}) should be noted:
\begin{enumerate}
\item
It exhibits the Regge type $x^{-\lambda}$ increase with decreasing $x$ where
the exponent $\lambda=\lambda_{BFKL}$ can have potentially large magnitude
$\approx 1/2$. The quantity $1+\lambda_{BFKL}$ is equal to the intercept
of the so called BFKL pomeron which corresponds to the hard QCD interactions.
Its potentially large magnitude ($\approx 1.5$) should be contrasted with
the intercept $\alpha_{soft} \approx 1.08$ of the effective "soft" pomeron
which has been determined from the phenomenological analysis of the high
energy behaviour of hadronic and photoproduction total cross-sections
\cite{DOLA}.
\item
It exhibits the $(Q_t^2)^{1/2}$ growth with increasing $Q_t^2$ modulated
by the Gaussian distribution in $ln(Q_t^2)$ of width increasing as
$ln^{1/2}(1/x)$ with decreasing $x$.  The Gaussian factor
reflects the diffusion pattern inherent in the BFKL equation.  The increase
of the function  $f(x,Q_t^2)$ as $(Q_t^2)^{1/2}$ is due to the fact that
the leading twist anomalous dimension is equal to $1/2$ for $\omega =
\lambda_{BFKL}$ (see eq. (\ref{onehalf})).
This shift of the anomalous dimension is the result of the (infinite) $LL1/x$
resummation.
\item
The diffusion pattern of the solution of the BFKL equation
is the direct consequence of the absence of transverse momentum ordering
along the gluon chain.  In this respect the BFKL dynamics is different
from that based on the (leading order) Altarelli-Parisi evolution which
corresponds to the strongly  ordered transverse momenta.
The interrelation between the diffusion of transverse momenta towards
both the infrared and ultraviolet regions {\bf and} the increase of gluon
distributions
with decreasing $x$ is an  important property of QCD at low $x$.
It has important consequences for the structure of the hadronic final
state in deep inelastic scattering at small $x$.\\
\end{enumerate}

The structure functions $F_{2,L}(x,Q^2)$ are  described  at small $x$
by the
diagram of Fig.5 which gives the following relation between
the structure functions and the unintegrated distribution $f$:
\begin{equation}
F_{2,L}(x,Q^2)=\int_x^1{dx^{\prime}\over x^{\prime}}\int
{dQ_t^2\over Q_t^2}F^{box}_{2,L}(
x^{\prime},Q_t^2,Q^2)f({x\over x^{\prime}},Q_t^2).
\label{ktfac}
\end{equation}
The functions  $F^{box}_{2,L}(x^{\prime},Q_t^2,Q^2)$ may be regarded as  the
structure
functions of the off-shell gluons with  virtuality
$Q_t^2$.
They are described by the quark box (and crossed box) diagram contributions
to the
photon-gluon interaction in the upper part of the diagram of Fig. 5.
The small $x$ behaviour of the structure functions reflects the small
$z$ ($z = x/x^{\prime}$) behaviour of the gluon distribution $f(z,Q_t^2)$.\\

The equation (\ref{ktfac}) is an example of the "$k_t$ factorization theorem"
which relates the measurable quantities (like DIS structure functions) to
the convolution in both longitudinal as well as in transverse momenta of the
universal gluon distribution $f(z,Q_t^2)$ with the cross-section
(or structure function) describing the interaction of the "off-shell" gluon
with the hard probe \cite{CIAFKT,KTFAC}.
The $k_t$ factorization theorem is the basic tool for
calculating the observable quantities in the small $x$ region in terms of the
(unintegrated) gluon distribution $f$ which is the solution of the BFKL
equation.\\

The leading - twist part of the $k_t$ factorization formula can be rewritten
in a collinear factorization form.  The leading small $x$ effects are then
automatically resummed in the
 splitting functions and in the coefficient functions. The $k_t$
factorization theorem   can in fact be used as the tool for calculating
these quantities.   We shall demonstrate this using the example of the
$P_{qg}$ splitting function which is responsible for the evolution
of the quark densities at low $x$ (see eqs.(\ref{apg}) and
(\ref{qdbg})).\\

{}From the $k_t$ factorization theorem we get:
\begin{equation}
Q^2{d \bar F_2(\omega,Q^2)\over d Q^2} =
\int
{dQ_t^2\over Q_t^2}Q^2{d \bar F^{box}_{2}(
\omega, Q_t^2,Q^2)\over d Q^2}\bar f(\omega,Q_t^2)
\label{ktmom}
\end{equation}
where $ \bar F_2(\omega,Q^2)$ and $\bar F^{box}_{2}(
\omega, Q_t^2,Q^2)$ are the corresponding  moment functions i.e.
\begin{equation}
 \bar F_2(\omega,Q^2)=\int_0^1{dx\over x}x^{\omega}F_2(x,Q^2)
\label{f2mom}
\end{equation}
and
\begin{equation}
\bar F^{box}_{2}(
\omega, Q_t^2,Q^2)=\int_0^1{dx\over x}x^{\omega}
 F^{box}_{2}(x, Q_t^2,Q^2).
\label{fboxmom}
\end{equation}
Inserting the (inverse) Mellin representation of $\bar F^{box}_{2}(
\omega, Q_t^2,Q^2)$
\begin{equation}
\bar F^{box}_{2}(
\omega, Q_t^2,Q^2)={1\over 2 \pi i} \int_{1/2-i\infty}^{1/2+i\infty}
d\gamma \tilde F^{box}_{2}(
\omega,\gamma)\left(Q^2\over Q_t^2 \right)^{\gamma}
\label{mbox}
\end{equation}
and of $\bar f(\omega,Q_t^2)$ (see eq.(\ref{imt})) into eq.(\ref{ktmom})
we get the following representation of
$Q^2 {d \bar F_2(\omega,Q^2)\over d Q^2}$
\begin{equation}
Q^2 {d \bar F_2(\omega,Q^2)\over d Q^2}=
{1\over 2 \pi i} \int_{1/2-i\infty}^{1/2+i\infty}
d\gamma \gamma \tilde F^{box}_{2}(\omega,\gamma) \tilde f(\omega,\gamma)
(Q^2)^{\gamma}
\label{ktdgamma}
\end{equation}
The leading twist part of the integral in eq. (\ref{ktdgamma})
is controlled by the anomalous dimension
$\gamma_{gg}({\bar \alpha_s\over \omega})$ which is a pole of
$\tilde f(\omega,\gamma)$ in the complex $\gamma$ plane.
It gives the following contribution to
 $Q^2{d \bar F_2(\omega,Q^2)\over d Q^2}$:
\begin{equation}
Q^2 {d \bar F_2(\omega,Q^2)\over d Q^2}=
 \gamma_{gg}^2({\bar \alpha_s\over \omega})
\tilde F^{box}_{2}\left(\omega,
\gamma= \gamma_{gg}({\bar \alpha_s\over \omega})\right)
g(\omega,Q^2)
\label{coll1}
\end{equation}
where we have taken into account equations (\ref{adbeh}), (\ref{intgm})
and (\ref{g0}).
In the so - called DIS scheme  the QCD - improved parton model relation
(\ref{f2pm}) holds beyond the leading order and so
we have the following relation between
 $Q^2{d \bar F_2(\omega,Q^2)\over d Q^2}$ and $g(\omega,Q^2)$
\begin{equation}
Q^2 {d \bar F_2(\omega,Q^2)\over d Q^2}=2\sum_i e_i^2 P_{qg}(\omega,\alpha_s)
g(\omega,Q^2) +....
\label{coll2}
\end{equation}
where  $P_{qg}(\omega,\alpha_s)$ is the moment of the splitting function
$P_{qg}(z,\alpha_s)$.  Comparing eq.(\ref{coll2}) with (\ref{coll1}) we get the
following prescription for $P_{qg}(\omega,\alpha_s)$ in the leading $ln(1/x)$
(or rather in the leading $1/\omega$) approximation \cite{KTFAC,MK}:
\begin{equation}
P_{qg}(\omega,\alpha_s)=
{\gamma_{gg}^{2}({\bar \alpha_s\over \omega})\tilde F^{box}_{2}
\left(\omega=0,\gamma= \gamma_{gg}({\bar \alpha_s\over \omega})\right)
\over 2\sum_i e_i^2 }
\label{pqgf}
\end{equation}
The function $\gamma^{2}\tilde F^{box}_{2}
(\omega=0,\gamma)$ can be expanded into  the following power series
in $\gamma$
\begin{equation}
\gamma^{2}\tilde F^{box}_{2}
(\omega=0,\gamma)=\bar \alpha_s( d_0 + \sum_{n=1}^{\infty}
d_n \gamma^n )
\label{pser1}
\end{equation}
It should be noted that the function $\tilde F^{box}_{2}
(\omega=0,\gamma)$ has the $1/\gamma^2$ singularity at $\gamma=0$.
This follows from the fact that $\bar F^{box}_{2}
(\omega, Q_t^2, Q^2) \sim ln(Q^2/Q_t^2)$ at large $Q^2/Q_t^2$ because
of the collinear singularity. The function $\gamma^2
\tilde F^{box}_{2}(\omega=0,\gamma)$ is regular at $\gamma=0$ and can be
expanded in the power series (\ref{pser1}).\\

{}From the power series (\ref{pser1}) we get the following expansion:
\begin{equation}
\gamma_{gg}^{2}({\bar \alpha_s\over \omega})\tilde F^{box}_{2}
\left(\omega=0,\gamma= \gamma_{gg}({\bar \alpha_s\over \omega})\right)=
\bar \alpha_s \left [ d_0 + \sum_{n=1}^{\infty}
d_n \gamma_{gg}^n({\bar \alpha_s\over \omega})\right]
\label{prs2}
\end{equation}
Combining this expansion with the expansion (\ref{adexp}) of the anomalous
dimension $\gamma_{gg}({\bar \alpha_s\over \omega})$  we get the following
expansion of the splitting function $P_{qg}(z,\alpha_s)$ at small $z$:
\begin{equation}
zP_{qg}(z,\alpha_s)={\alpha_s\over 2 \pi}zP^{(0)}(z) +
(\bar \alpha_s)^2 \sum_{n=1}^{\infty}b_n
{[\bar \alpha_s ln(1/z)]^{n-1}\over (n-1)!}
\label{zpqgf}
\end{equation}
where the coefficients $b_n$ can be expressed in terms of $c_n$
and  $d_n$ appearing in the expansions (\ref{adexp}) and (\ref{pser1})
respectively.  The first term on the right hand side of eq. (\ref{zpqgf})
vanishes at $z=0$.  It should be noted that
 the  splitting function $P_{qg}$ when compared with the splitting
function $P_{gg}$
is formally non-leading at small $z$.
For moderately small values of $z$ however,
when the first few terms in the expansions (\ref{adexp}) and (\ref{zpqgf})
dominate, the BFKL effects can be much more important
in $P_{qg}$  than in $P_{gg}$.
This comes from the fact that in the expansion (\ref{zpqgf})
all coefficients $b_n$ are different from zero while in eq. (\ref{adexp})
we have $c_2=c_3=0$ \cite{JAR}.
The small $x$ resummation efects within the conventional QCD evolution
formalism have recently been discussed in refs. \cite{EKL,HBRW,BFORTE,FRT}. \\

A more general treatment of the gluon ladder than that which follows
from the BFKL formalism is  provided by
the CCFM equation based on angular ordering along the gluon chain
\cite{CCFM,KMS1}.
This equation embodies both the BFKL equation at small $x$ and the
conventional Altarelli-Parisi evolution at large $x$.
The unintegrated gluon distribution $f$ now acquires
dependence upon an additional scale $Q$
which specifies the maximal angle of gluon
emission.
The CCFM equation has a form analogous to that of the "folded" BFKL equation
(\ref{bfklr}):
$$
f(x,Q_t^2,Q^2)=\hat f^0(x,Q_t^2,Q^2)+ $$
\begin{equation}
\bar \alpha_s
\int_x^1{dx^{\prime}\over
x^{\prime}} \int {d^2 q\over \pi q^2} \Theta
(Q-qx/x^{\prime})\Delta_R({x\over x^{\prime}},Q_t^2,q^2)
{Q_t^2 \over (\mbox{\boldmath $q$}+
\mbox{\boldmath $Q_t$})^2}
f(x^{\prime},(\mbox{\boldmath $q$}+
\mbox{\boldmath $Q_t$})^2,q^2))
\label{ccfm}
\end{equation}
where the theta function $\Theta(Q-qx/x^{\prime})$ reflects the angular
ordering constraint on the emitted gluon.
The "non-Sudakov" form-factor $\Delta_R
(z,Q_t^2,,q^2  )$ is now given by the following formula:
\begin{equation}
\Delta_R(z,Q_t^2,q^2)=exp\left[-\bar \alpha_s\int_z^1 {dz^{\prime}
\over z^{\prime}} \int {dq^{\prime 2}
\over q^{\prime 2}}\Theta (q^{\prime 2}-(qz^{\prime})^2)
\Theta (Q_t^2-q^{\prime 2})\right]
\label{ns}
\end{equation}
Eq.(\ref{ccfm}) still contains only the singular term of the
$g \rightarrow gg$ splitting function
at small $z$. Its generalization which would
include
remaining parts of this vertex (as well as quarks) is possible.\\

In Fig. 6 we show the results for the structure function $F_2$ calculated
from the $k_t$ factorization theorem using the function $f$ obtained from
the CCFM equation \cite{CCFMF2}.
We confront these predictions with the most recent data
from the H1 and ZEUS collaborations at HERA \cite{H1,ZEUS} as well as
with the results
of the analysis which was based on the Altarelli-Parisi equation alone
without the small $x$ resummation effects being included in the formalism
\cite{MRS,GRV}.
In the latter case the singular small $x$ behaviour of the gluon
and sea quark distributions
 has to be introduced in a parametrization of the starting
distributions at the moderately large reference scale $Q^2=Q_0^2$
 (i.e. $Q_0^2 \approx 4 GeV^2$ or so) \cite{MRS}.  One can also
generate steep behaviour dynamically starting from the
non-singular "valence-like" parton distributions at some very low
scale $Q_0^2=0.35GeV^2$ \cite{GRV}. In the latter case the gluon and sea
quark
distributions exhibit  "double logarithmic behaviour" (\ref{dl}).
For very small values of the scale $Q_0^2$ the evolution lenghth $\xi(Q^2,
Q_0^2)$
can become large for moderate and large values of $Q^2$ and the "double
logarithmic" behaviour (\ref{dl}) is, within the limited region of $x$,
similar to that corresponding to the power like increase of the type
$x^{-\lambda}$, $\lambda \approx 0.3$.  This explains similarity between
the theoretical curves presented in Fig.6.  The theoretical results
also show  that an inclusive quantity like $F_2$ is not the
best discriminator for revealing the dynamical details at low $x$.
One may however hope that this can be provided by studying the structure
of the hadronic final state in deep inelastic scattering and this possibility
will be briefly discussed in the next section.

\section*{5. The structure of the hadronic final state in deep inelastic
scattering at low $x$}
It is expected that absence of transverse momentum ordering along the gluon
chain which leads to the correlation between the increase of the
structure function  with  decreasing $x$ and the diffusion
of transverse momentum should reflect itself in the behaviour of
less inclusive
quantities than the structure function $F_2(x,Q^2)$.  The dedicated
measurements of the low $x$ physics which are particularly sensitive
to this correlation are the deep inelastic plus
jet events, transverse energy flow
in deep inelastic scattering, production of jets separated by the large
rapidity gap and dijet production in deep inelastic scattering.
The diagrammatic illustration of these measurements is presented in
Fig. 7.\\

In principle  deep inelastic lepton scattering containing a measured jet
can provide a very clear test of the BFKL dynamics at low $x$
\cite{MJET,BJET,KMSJET}.
The idea is to study deep inelastic ($x,Q^2$) events which
contain an identified jet ($x_j,k_{Tj}^2$) where
$x<<x_j$ and $Q^2 \approx k_{Tj}^2$.  Since we choose events with
$Q^2 \approx k_{Tj}^2$ the leading order QCD evolution  (from $k_{Tj}^2$ to
$Q^2$) is neutralized and attention is focussed on the small $x$, or rather
small $x/x_j$ behaviour.  The small $x/x_j$ behaviour
of jet production is generated  by the gluon radiation as shown
in the diagram of Fig. 7a.  Choosing the
configuration $Q^2 \approx k_{Tj}^2$ we eliminate by definition
gluon emission which corresponds to strongly ordered transverse momenta
i.e.  that emission which is responsible for the LO QCD evolution.
The measurement of jet production in this configuration may therefore test
more directly the $(x/x_j)^{-\lambda}$ behaviour which is generated
by the BFKL equation where the transverse momenta are not ordered.
The recent H1 results concerning  deep inelastic plus jest events
are consistent with the increase of the cross-section with decreasing
$x$ as predicted by the BFKL dynamics {\cite{H1JET}. \\

Conceptually similar process is that of the two-jet production
separated by a large rapidity gap $\Delta y$ in hadronic
collisions or in photoproduction as illustrated in Fig. 7c
\cite{DDUCA,JAMESJ}.
Besides the characteristic $exp(\lambda \Delta y)$ dependence
of the two-jet cross-section one expects significant
weakening of the azimuthal back-to-back correlations
of the two jets.  This is the direct consequence of the
absence of transverse momentum ordering along the gluon
chain in the diagram of Fig. 7c.\\

Another measurement which should be sensitive to the QCD
dynamics at small $x$ is that of the transverse energy flow in deep inelastic
lepton scattering in the central region away from the
current jet and from the proton remnant as illustrated in Fig. 7b.
\cite{KMSET}.
The  BFKL dynamics predicts
in this case a substantial amount of transverse energy  which should
increase with decreasing $x$.  The experimental data
are consistent with this theoretical expectation \cite{H1JET}.
\\
Absence of transverse momentum ordering  also implies weakening of the
back-to-back azimuthal correlation of dijets produced close
to the photon fragmentation region (see Fig. 7d) \cite{DICKJET,AGKM}. \\

Another important process which is sensitive to the small $x$ dynamics
is  the deep inelastic diffraction \cite{ZDIF,H1DIF}.
Deep inelastic diffraction in $ep$ inelastic scattering is a process:
\begin{equation}
e(p_e)+p(p) \rightarrow e^{\prime}(p_e^{\prime}) + X +p^{\prime}(p^{\prime})
\label{disdif}
\end{equation}
where there is a large rapidity gap between the recoil proton
(or excited proton) and the hadronic system $X$ (see Fig. 8a).
To be precise
process (\ref{disdif}) reflects the diffractive disssociation
of the virtual photon.  Diffractive dissociation is described by the following
kinematical variables:
\begin{equation}
\beta={Q^2\over 2 (p-p^{\prime})q}
\end{equation}
\begin{equation}
x_P={x\over \beta}
\end{equation}
 \begin{equation}
t= (p-p^{\prime})^2.
\label{difv}
\end{equation}
Assuming that  diffraction dissociation is dominated by the pomeron
exchange as shown in Fig. 8b and that the pomeron is described by
a Regge pole one gets the following factorizable expression for the
diffractive structure function \cite{DOLADIF,ORSAYDIF,CTEQD,JAMESD,KGBK}:
\begin{equation}
{\partial F_2^{diff}\over \partial x_P \partial t}= f(x_P,t)F_2^P(\beta,Q^2,t)
\label{difsf}
\end{equation}
where the "flux factor" $f(x_P,t)$ is given by the following formula
:
\begin{equation}
f(x_P,t)=N{B^2(t)\over 16\pi} x_P^{1-2\alpha_P(t)}
\label{flux}
\end{equation}
with $B(t)$ describing the pomeron coupling to a proton and $N$ being the
normalisation factor.  The function $F_2^P(\beta,Q^2,t)$
is the pomeron structure function which in the (QCD improved) parton model
is related in a standard way to the quark and antiquark distribution
functions in a pomeron (see Fig. 9).
\begin{equation}
F_2^P(\beta,Q^2,t)=\beta \sum e_i^2[q_i^P(\beta,Q^2,t)+ \bar
q_i^P(\beta,Q^2,t)]
\label{f2pom}
\end{equation}
with $q_i^P(\beta,Q^2,t)=\bar q_i^P(\beta,Q^2,t)$.  The variable
$\beta$ which is the Bjorken scaling variable appropriate for
deep inelastic lepton-pomeron "scattering",  has the meaning of the
momentum fraction of the pomeron carried by the
probed quark (antiquark).  The quark distributions in a pomeron
are assummed to obey the standard Altarelli-Parisi evolution equations:
\begin{equation}
Q^2{\partial q^P\over \partial Q^2}=P_{qq} \otimes q^P + P_{qg} \otimes g^P
\label{app}
\end{equation}
with a similar equation for the evolution of the gluon distribution
in a pomeron.  The first
term on the right hand side of the eq. (\ref{app}) becomes negative
at large $\beta$,  while the second term remains positive and
is usually very small at large $\beta$ unless the gluon distributions
are large and have a hard spectrum.\\

In Figs. 10a,b we show  the theoretical results for the quantity
\begin{equation}
 F_2^D(\beta,Q^2)=\int_{x_{PL}}^{x_{PH}}dx_P \int_{-\infty}^0 dt
  {\partial F_2^{diff}\over \partial x_P \partial t}
\label{f2d}
\end{equation}
with $x_{PL}=0.0003$ and $x_{PH}=0.05$ based on the "conventional"
parton distributions in a pomeron vanishing as $(1-\beta)$ at $\beta=1$
and compare these results with the experimental data from HERA \cite{H1DIFP}.
 The data suggest that the slope
of $F_2^P$ as the function of $Q^2$ does not change sign even
at relatively large values of $\beta$.  This favours the hard
gluon spectrum in a pomeron \cite{ABRAMOWICZ,DAINTON}, and
should be contrasted with the behaviour of the structure function
of the proton which, at large $x$, decreases with increasing $Q^2$.
The data on inclusive diffractive production  favour the soft pomeron
with relatively low intercept. This is illustrated in
Fig. 11 where we plot the theoretical prediction for the quantity
\begin{equation}
F_2^{D(3)}(x_P,\beta,Q^2)=\int_{-\infty}^0dt
{\partial F_2^{diff}\over \partial x_P \partial t}
\label{f2d3}
\end{equation}
based on the "soft" pomeron with low intercept $\alpha_{soft}=1.1$
and compare these
predictions with the experimental data from HERA \cite{H1DIFP}.
 The diffractive production of vector mesons
 seems to require a "hard" pomeron contribution
\cite{DONNACHIE,WHITMORE,KENYON} . It has also
been pointed out  that the factorization property
(\ref{difsf}) may
not hold in models based
entirely on perturbative QCD when the pomeron is represented
by the BFKL ladder \cite{KOLYA,BARTELSD}.
There exist also models of deep inelastic diffraction which do not
rely on the pomeron exchange picture \cite{BUCHMD,EIR}.\\

\section*{6. Summary and conclusions}

In this lecture we have briefly described the QCD expectations
for deep inelastic lepton scattering at low $x$ which  follow from the
BFKL dynamics.  It leads to the indefinite increase of gluon distributions
with decreasing $x$ which is correlated with the diffusion of
 transverse momenta. This increase of gluon distribution
implies a similar increase of the structure functions through the
$g \rightarrow q \bar q$ transitions.
Besides discussing the theoretical and
phenomenological issues related to the description of the structure
function $F_2$ at low $x$ we have also emphasised
the role of studying the hadronic final state in deep inelastic scattering.
\\

The indefinite growth of parton distributions cannot go on forever
and has to be eventually stopped by parton screening which leads
to the parton saturation.  Most probably however
the saturation limit is still irrelevant for the small $x$ region
which is now being probed at HERA.
\\

We have limited ourselves to the large $Q^2$ region where perturbative
QCD is expected  to be applicable.  Specific problems of the low $
Q^2$, low $x$ region are discussed in ref. \cite{BBJK}.
Finally let us point out that the change of the dynamics with
the relevant scale is clearly visible in the data (see Fig. 2) and its
satisfactory explanation is perhaps one of the most challenging
problems of to-day.
\medskip\medskip\medskip

\section*{Acknowledgments}
I thank the  organizers of the Jubillee XXXVth Cracow School of
Theoretical
Physics for organizing an excellent meeting.
I thank Barbara Bade\l{}ek, Krzysztof Golec-Biernat,
Alan Martin and Peter Sutton for most enjoyable research collaborations
on the problems presented in this lecture.
I am grateful to Grey College and to Physics
Department of the University of Durham for their warm hospitality.
This research has been supported in part by
 the Polish State Committee for Scientific Research grant 2 P302 062 04 and
the EU under contracts n0. CHRX-CT92-0004/CT93-357.\\

{\Large {\bf Figure captions}}
\begin{enumerate}
\item
Kinematics of inelastic lepton-nucleon scattering in the one photon
exchange approximation and its relation through the optical theorem to
Compton scattering of the virtual photon. $p$ and $q$ denote
the four momenta of the nucleon and of the virtual photon resspectively.
\item
The total virtual photon cross-section plotted as a function of $W^2$
for different values of the photon virtuality $Q^2$.  The curves
correspond to the theoretical parametrizations \cite{GRV,DOLAF2}.
The figure is taken from ref. \cite{ALEVY}.
\item
The hand-bag diagram for the virtual Compton scattering on a nucleon; $k$
denotes
the four momentum of the struck quark (antiquark).  At high $Q^2$ and in the
infinite momentum frame of the nucleon $k \approx xp$
where $x$ is the Bjorken scaling variable.
\item
Diagrammatic representation of the BFKL equation (\ref{bfklr}).
\item
Diagrammatic representation of the $k_t$ factorization formula (\ref{ktfac}).
\item
A comparison of the HERA measurements of $F_2$ \cite{H1,ZEUS} with
the predictions based on the $k_t$ factorization formula (\ref{ktfac})
using for the unintegrated gluon distributions $f$ the solutions of the
CCFM equation (\ref{ccfm}) (continuous curve) and of the
approximate form of this
equation corresponding to setting
$\Theta(Q-q)$ in place of $\Theta(Q-qx/x^{\prime})$ and
$\Delta_R=1$ (dotted curve).
 We also show the values of $F_2$ obtained from
collinear factorization using the MRS(A$^{\prime})$ \cite{MRS} and
GRV \cite{GRV} partons (the figure is taken from  ref. \cite{CCFMF2}).
\item
Diagrammatic representation of the processes testing the BFKL
dynamics. (a) Deep inelastic scattering with the forward jet.
(b) $E_T$ flow in deep inelastic scattering. (c) Production
of jets separated by the large rapidity gap $\Delta y$. (d)
Dijet production in deep inelastic scattering (the figure is taken from
ref. \cite{ADMBLOIS}).
\item
(a) Kinematics of the large rapidity gap process $e + p \rightarrow
e^{\prime} + X +p^{\prime}$.  The wavy line represents the virtual photon.
(b) The pomeron exchange diagram for diffractive production of the
hadronic system $X$ by a virtual photon.  The wavy and zigzag lines
represent the virtual photon and pomeron respectively.
\item
The hand-bag diagram for the virtual Compton diffractive production.
The wavy and zigzag lines
represent the virtual photon and pomeron respectively.
\item
Theoretical predictions \cite{KGBK,KGBKP} for the diffractive structure
function $F_2^D(\beta,
Q^2)$ defined by eq. (\ref{f2d})
and their comparison with the data from HERA \cite{H1DIFP}.  The structure
function $F_2^D(\beta,
Q^2)$ is plotted (a) as the function of $Q^2$ for fixed values of $\beta$ and
(b) as the function of $\beta$ for fixed values of $Q^2$.  The figure is taken
from ref. \cite{KGBK}.
\item
Theoretical predictions based on the soft pomeron
exchange with low intercept $\alpha_{soft}=1.1$
\cite{KGBK,KGBKP} for the diffractive structure function $F_2^{D(3)}
(x_P,\beta,
Q^2)$ defined by eq. (\ref{f2d3})
and their comparison with the data from HERA \cite{H1DIFP}.
 The structure function $F_2^{D(3)}
(x_P,\beta,
Q^2)$ is plotted as the function of $x_p$ for different values of
$\beta$ and $Q^2$.
\end{enumerate}
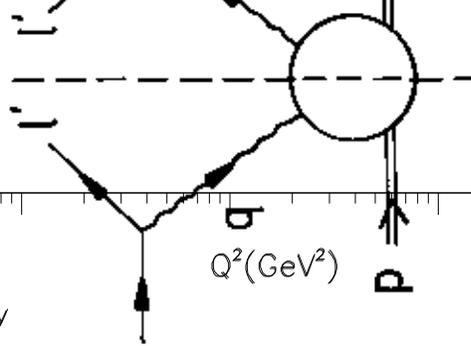
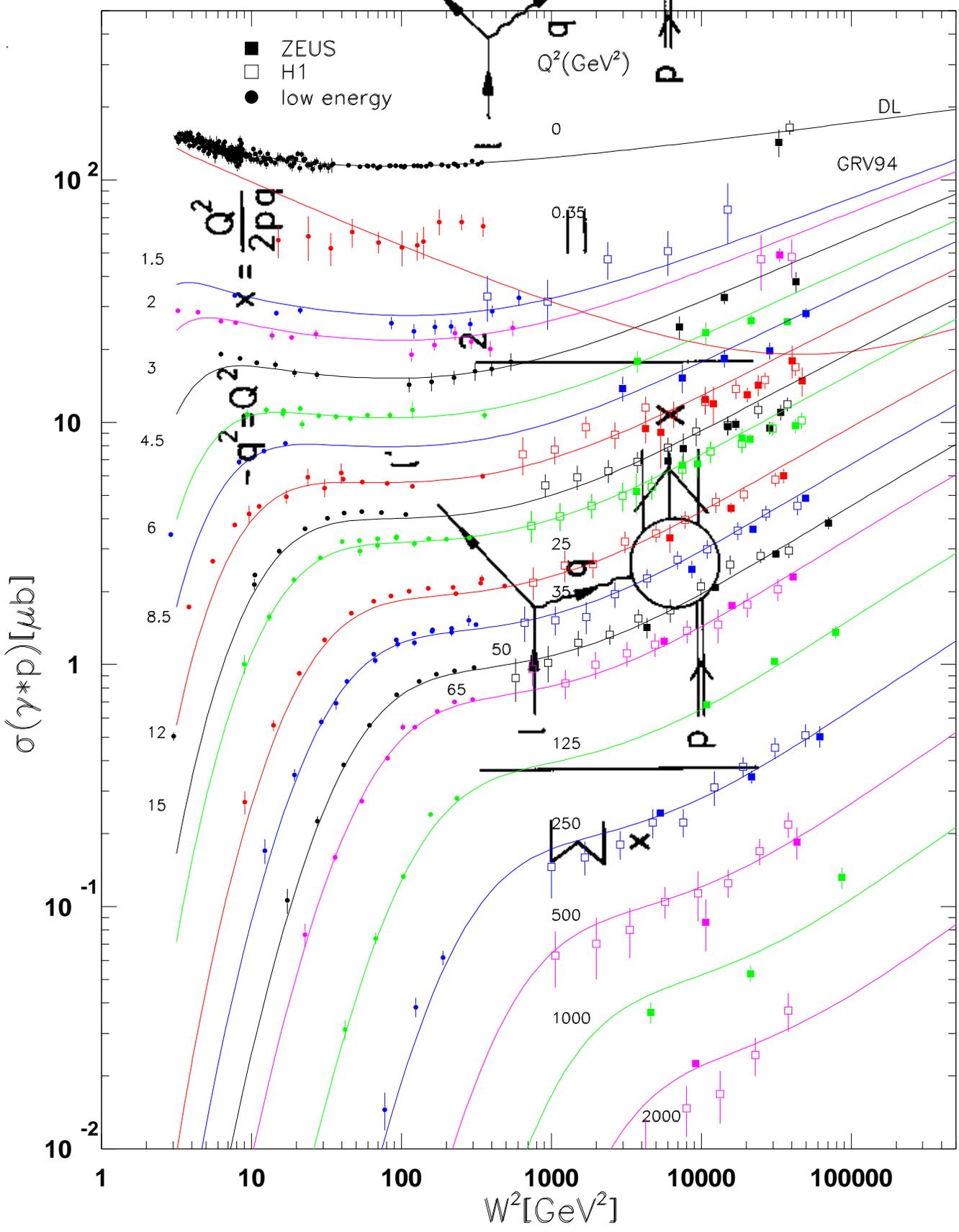
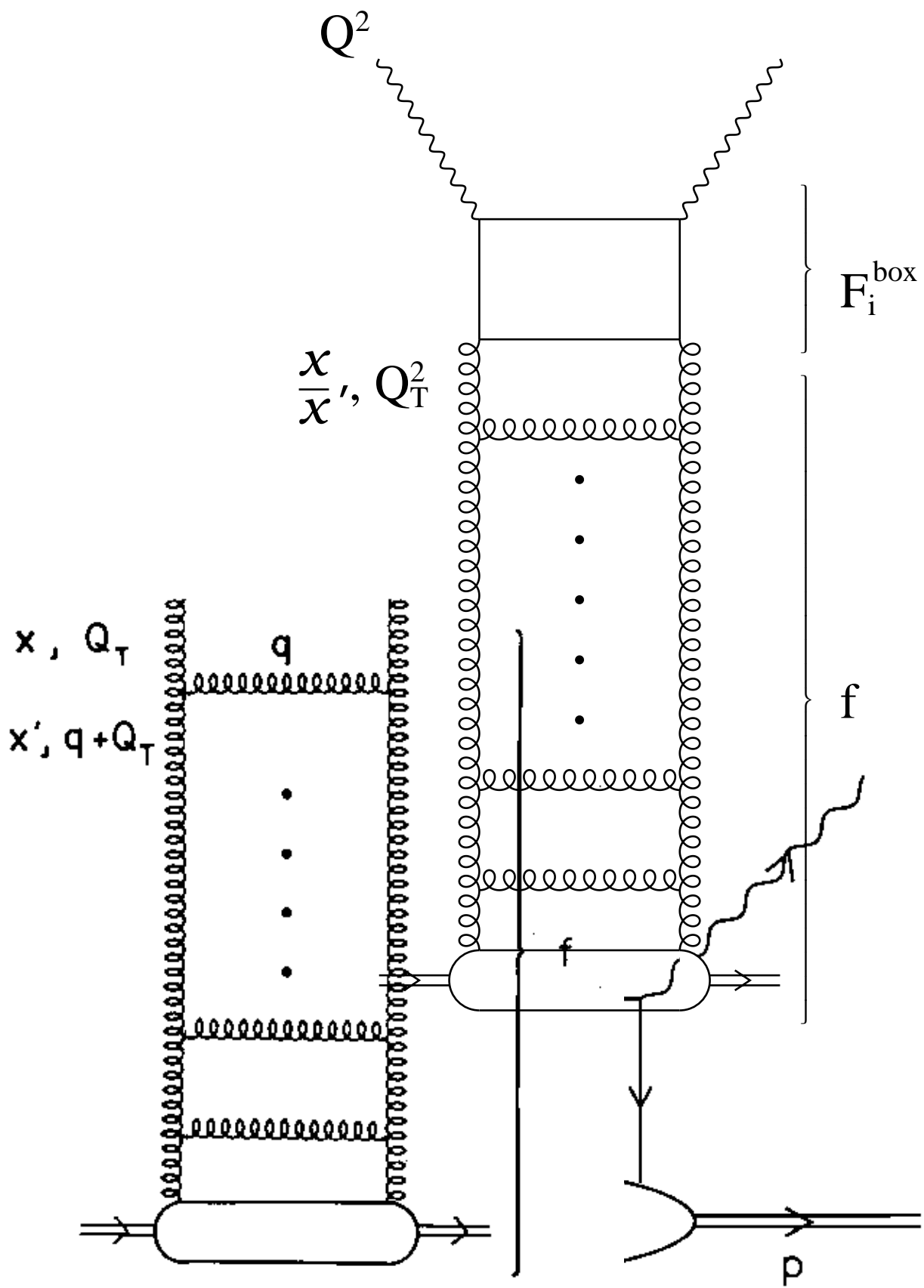
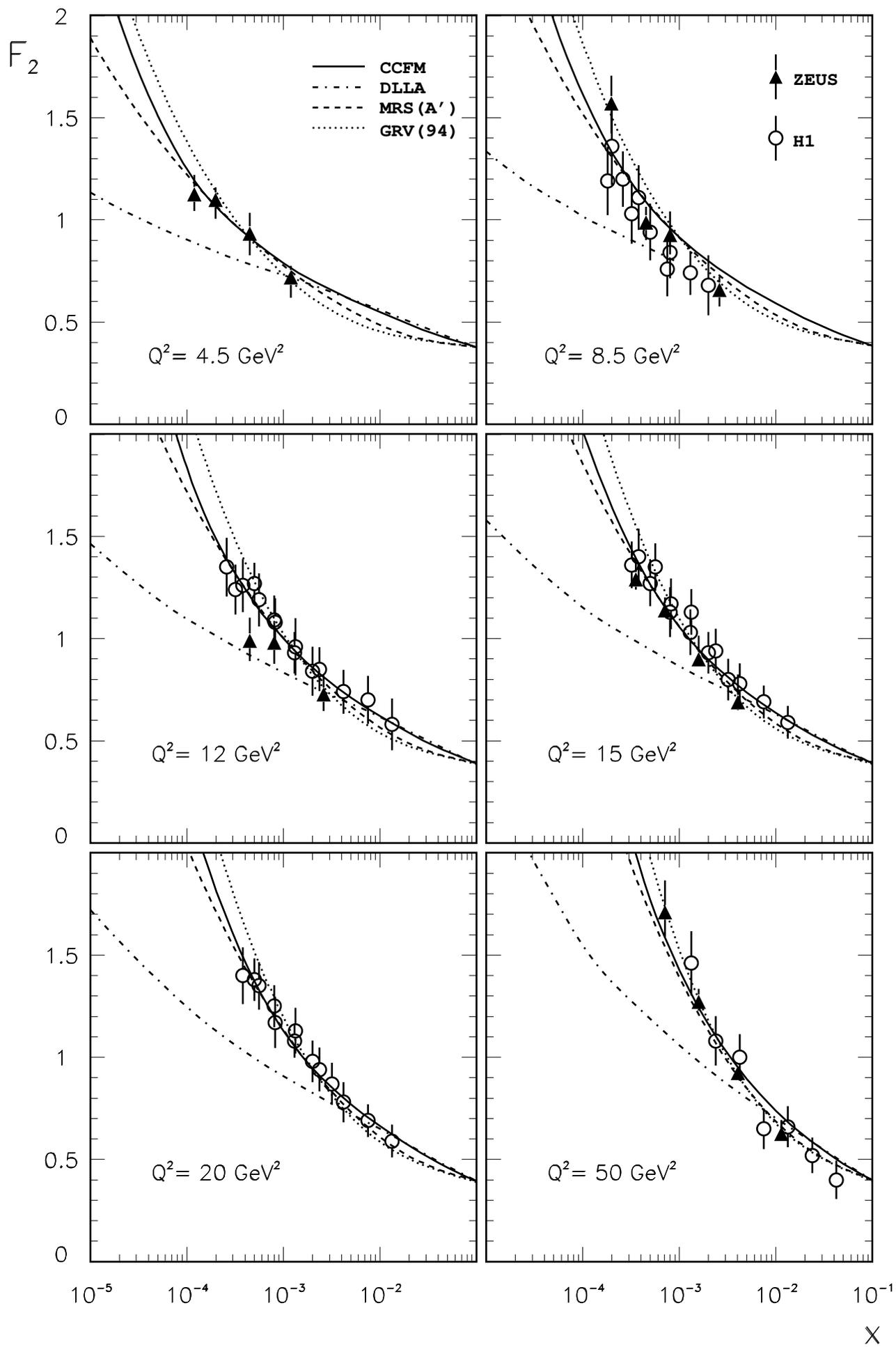
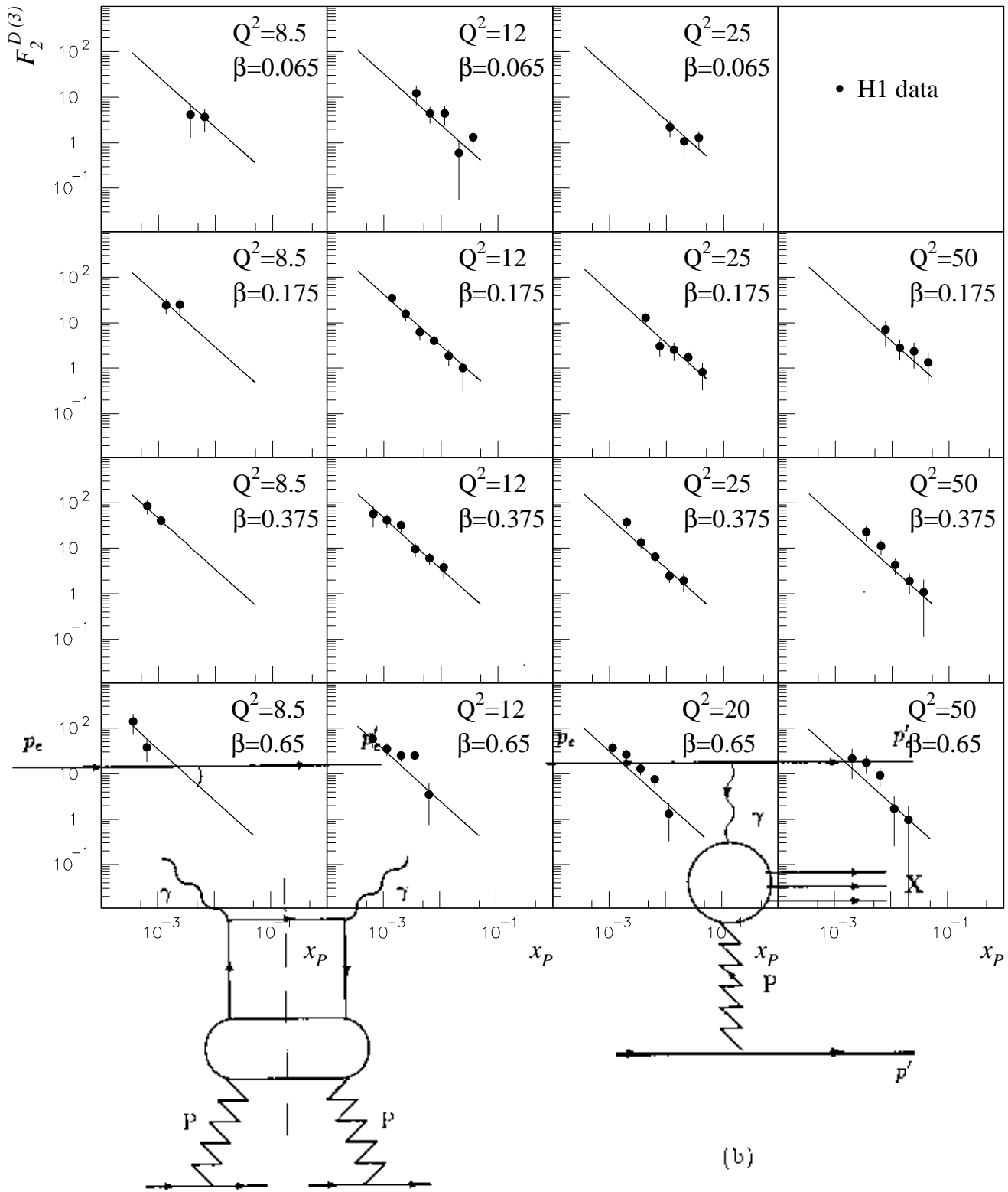
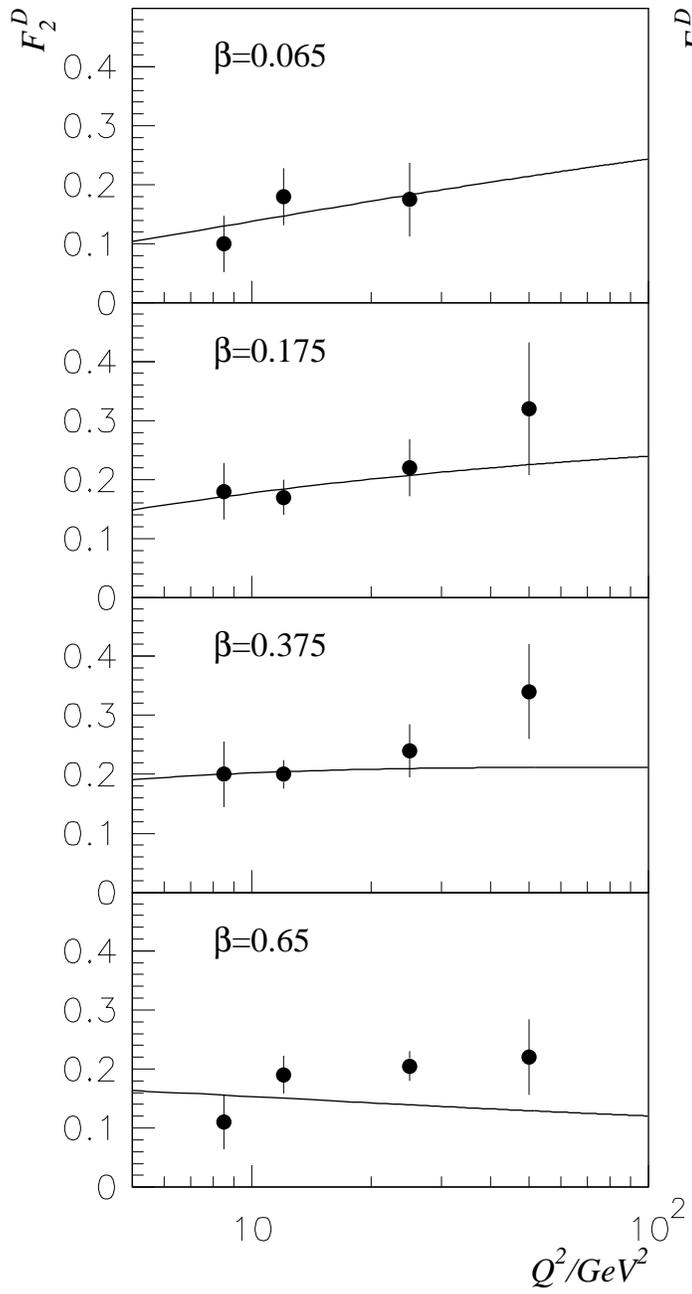
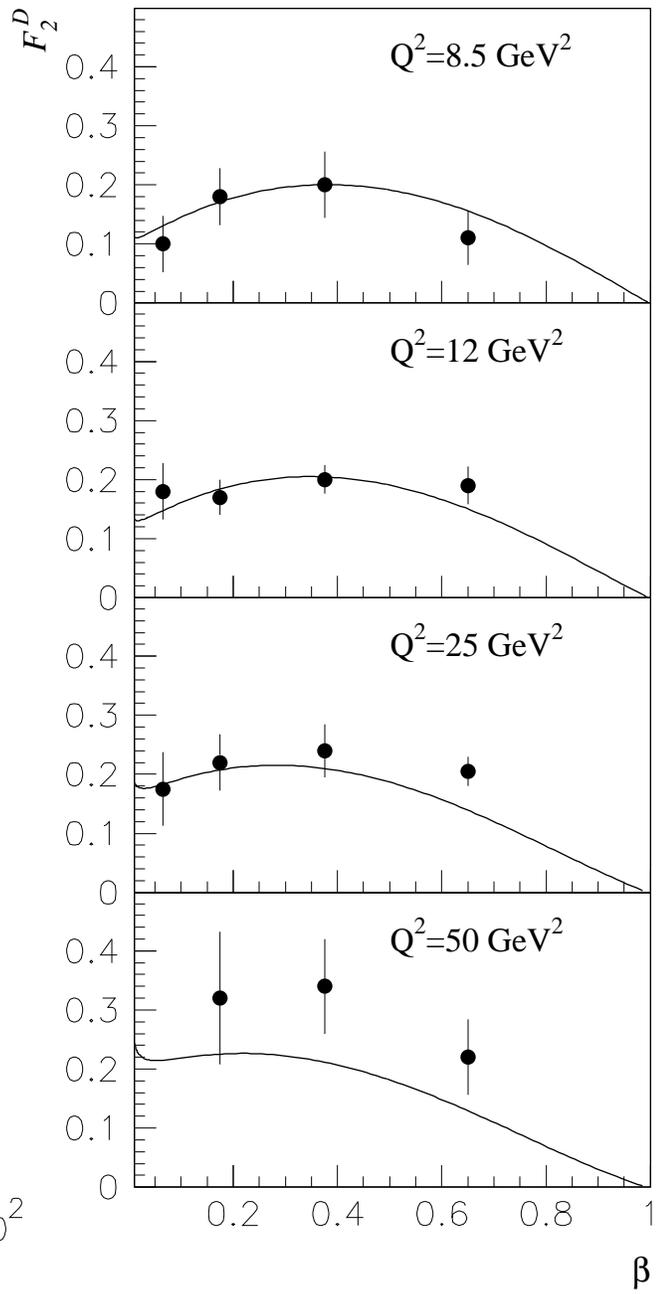
\end{document}